\documentclass[]{jfm}
\usepackage{graphicx}
\usepackage{newtxtext}
\usepackage{newtxmath}
\usepackage{natbib}
\usepackage{hyperref}
\hypersetup{
    colorlinks = true,
    urlcolor   = blue,
    citecolor  = black,
}

\newcommand{\RomanNumeralCaps}[1]

\title{Impulse-driven capillary detachment}

\author{
Dilip Kr.\ Maity\aff{1}
  \corresp{\email{dilmaity@gmail.com}},
Sandip Dighe\aff{1},
Nilamani Sahoo\aff{1}
\and
Tadd Truscott\aff{1}
  \corresp{\email{taddtruscott@gmail.com}}
}

\affiliation{
\aff{1}Department of Mechanical Engineering, Physical Science and Engineering Division, King Abdullah University of Science and Technology, Thuwal 23955, Kingdom of Saudi Arabia
}

\begin{document}
\maketitle

\begin{abstract}
Capillary interfaces subjected to impulsive forcing arise in many natural and technological systems, yet the pathway by which rapid substrate motion is converted into droplet detachment remains unclear.
Here we study this process in a controlled setting: a liquid droplet resting on a taut wire that is plucked and suddenly released. 
The resulting transverse wave imparts a brief inertial forcing at the droplet base, initiating rapid stretching that precedes sheet formation and jet breakup. 
We show that the maximum extension prior to detachment is set by the mechanical work transmitted from the wire through capillary traction at the three-phase contact line, balanced by viscous dissipation during filament extension. 
This energetic balance
identifies the contact line as the pathway by which mechanical impulse is
converted into capillary deformation and governs impulsive droplet detachment.
\end{abstract}



\section{Introduction}
\label{sec:headings}

\textcolor{black}{
Capillary interfaces often experience rapid forcing when the supporting substrate undergoes sudden motion. 
In such events, a mechanical impulse originating in the solid must be transmitted to the liquid through the three-phase contact line, initiating  deformation and eventual
detachment. 
Although the relevant physical ingredients are well known (i.e., capillarity, inertia, and viscous dissipation), how they combine to determine droplet release under impulsive forcing has lacked a simple organizing principle.}

\begin{figure}
\centering
\includegraphics[width=1\textwidth]{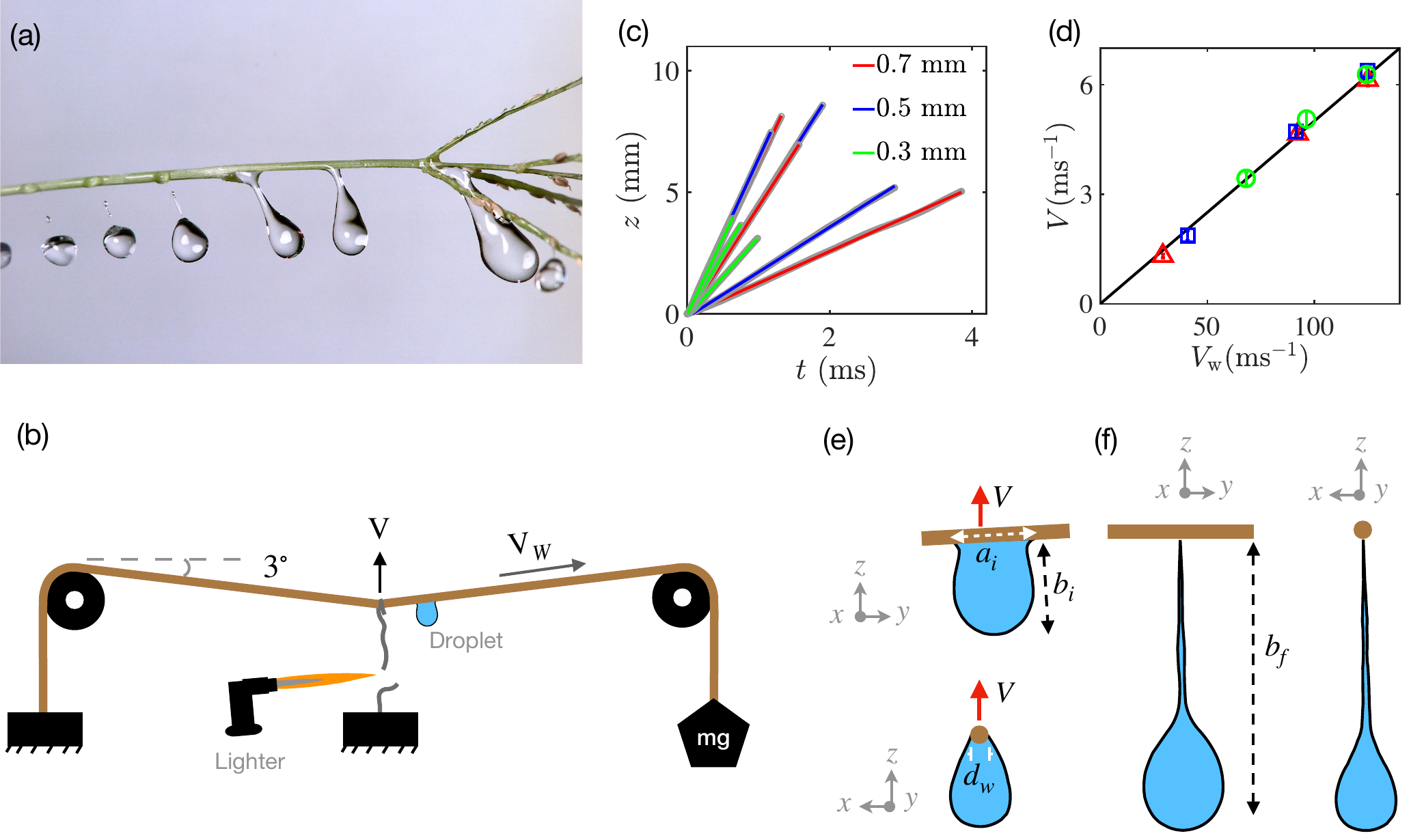}
\vspace{2mm}
\refstepcounter{figure}
\noindent\parbox{\textwidth}{%
{Figure \thefigure.}
{
(a) A droplet ejecting from a wire is analogous to a raindrop detaching from grass blades after a sudden motion (see supplemental video 1).  
(b) Schematic of the experimental setup showing a copper wire held in tension by a centrally attached nylon wire, which is pulled down and then suddenly released.  
(c) The vertical displacement ($z$) of the wire at the droplet position for three wire diameters and multiple tensions (2.94 to 53.96~N). (d) Vertical wire speed $V$ as a function of horizontal wave speed $V_{\mathrm{w}}$, line represents $V=\sqrt{T/M} \tan{\theta}$. (e) Equilibrium droplet resting on the wire before excitation contact length ($a_i$) and droplet height ($b_i$). (f) Droplet shape just before it is released from the wire with final height $b_f$.}}
\label{Fig:schematic}
\end{figure}

Droplets are often detached from surfaces by sudden motions, such as raindrops flung from grass (Fig.\ref{Fig:schematic}a),  water shaken off from bird feathers (\cite{ortega2012aerial}) or wet mammals (\cite{dickerson2012wet}), or fog-harvesting (\cite{bai2011controlled,ju2012multi}) fibers that vibrate in the wind or mechanical means (\cite{wang2025bioinspired}). The process of detachment usually involves sufficient acceleration for the liquid droplet to be separated from the solid. Early in the process a jet forms between the solid and droplet which eventually breaks apart fragmenting into many smaller droplets. The breakup of droplets under rapid forcing arises from the simultaneous interplay of inertia driving elongation, viscosity dissipating kinetic energy, and surface tension resisting deformation. These forces act in the droplet bulk, at the droplet-solid, and droplet-air interfaces over rapid timescales.

The jet fragmentation process is familiar 
in nature, such as raindrops breaking up in a storm (\cite{villermaux2009single}), and in technology, such as inkjet printing (\cite{lohse2022fundamental}).
Extensive theoretical and experimental studies (\cite{eggers1993universal,eggers1997nonlinear,wagner2003role, villermaux2004ligament, villermaux2009single,bhat2010formation, keshavarz2016ligament}) have explored these forces in classical contexts such as droplet pinch-off and filament thinning. These frameworks have proven especially powerful in revealing the structure of fluid necking and capillary instabilities. However, droplet breakup becomes more complex under sudden shear forces (between the solid-liquid interface), and recent work shows that acoustic, magnetic or vibrational excitation, such as selective withdrawal, magnetic field or substrate shaking, can change conditions and modes of breakup (\cite{vincent2014remnants, pan2020regime,farhan2018universal, ojaghlou2018dynamical,lejeune2019drop, gilet2025leaf}).

Here, we explore one of simplest liquid-solid experiments where 
a liquid droplet rests on a taut wire that is pulled downward in the middle and then suddenly released (Fig.~\ref{Fig:schematic}b), a process we refer to as plucking the wire. The resulting impulsive excitation launches a transverse wave along the wire, imparting localized vertical acceleration at the droplet base. This inertia input 
triggers a rapid stretching stage that precedes sheet formation, rim collision, and jet fragmentation. We resolve these transient stages and examine how the interplay of inertia, surface tension, and viscosity governs the breakup pathway through high-speed imaging. We show that the stretching preceding detachment is governed by a simple energetic balance: mechanical work transmitted from the wire through capillary traction at the contact line is balanced by viscous dissipation within the elongating filament. This balance predicts the maximum droplet extension and therefore the release time across a wide range of fluids and excitation conditions.

\section {Experimental method}
 A representative image of a droplet being ejected from a grass blade following an impulsive disturbance is shown in Fig.\ref{Fig:schematic}a, illustrating a natural phenomenon such as raindrop detachment caused by wind or mechanical agitation. To investigate this process systematically, we designed a laboratory setup using a taut wire under tension, as depicted in Fig.\ref{Fig:schematic}b. A 160\,mm long copper wire is horizontally suspended between two pulleys, with one end clamped to a fixed point and the other attached to a mass \( m \) to impose a controlled tension \( T = mg \). A nylon wire is fastened at the midpoint and pre-tensioned to form a small deflection angle ($3^\circ$). A droplet was deposited 1 cm away from the midpoint. We positioned the droplet close to the midpoint of the wire, just off the center, so that it primarily experienced vertical motion. Beyond this region, the wire exhibited a slight wavy motion. The nylon wire is burned by a lighter resulting in an impulsive upward motion of the wire due to the sudden release of tension. This method is more effective than cutting to reduce initial vibrations. We systematically vary  wire diameter (0.3, 0.5, \& 0.7~mm), droplet volume (ranging from 2 to 16~$\mu$L), and applied tension (ranging from 2.94 to 53.96~N). Experiments are conducted using water ($\mu = 0.96~\mathrm{mPa\,s}$, $\sigma = 72~\mathrm{mN\,m^{-1}}$, $\rho = 998~\mathrm{kg\,m^{-3}}$), a 72\% wt. glycerin--water mixture ($\mu = 25~\mathrm{mPa\,s}$, $\sigma = 65~\mathrm{mN\,m^{-1}}$, $\rho = 1185~\mathrm{kg\,m^{-3}}$), ethanol ($\mu = 1.1~\mathrm{mPa\,s}$, $\sigma = 22.1~\mathrm{mN\,m^{-1}}$, $\rho = 788~\mathrm{kg\,m^{-3}}$), and a 7 CMC aqueous SDS solution ($\mu = 0.96~\mathrm{mPa\,s}$, $\sigma = 33~\mathrm{mN\,m^{-1}}$, $\rho = 998~\mathrm{kg\,m^{-3}}$).

 The vertical displacement of the wire center over time, shown in Fig.\ref{Fig:schematic}c, confirms that the impulse causes a rapid rise with nearly constant velocity. This velocity $V$ is found to scale linearly with the transverse wave speed $V_{\mathrm{w}} = \sqrt{T/M}$, where $M$ is the mass per unit length of the wire, consistent with the relation $V = V_{\mathrm{w}} \tan\theta= 0.053 V_{\mathrm{w}}$ (Fig.\ref{Fig:schematic}d). 

Prior to excitation, the droplet rests stably on the wire as shown in Fig.\ref{Fig:schematic}e, characterized by contact length $a_i$ and vertical extent $b_i$. Following the impulsive pluck, the droplet stretches upward (Fig.\ref{Fig:schematic}f), forming a narrowing column of fluid that results in the formation of a jet with a final height just before detachment denoted as $b_f$.

\section{Results and discussion}
\begin{figure}
\centering
\includegraphics[width=0.9\textwidth]{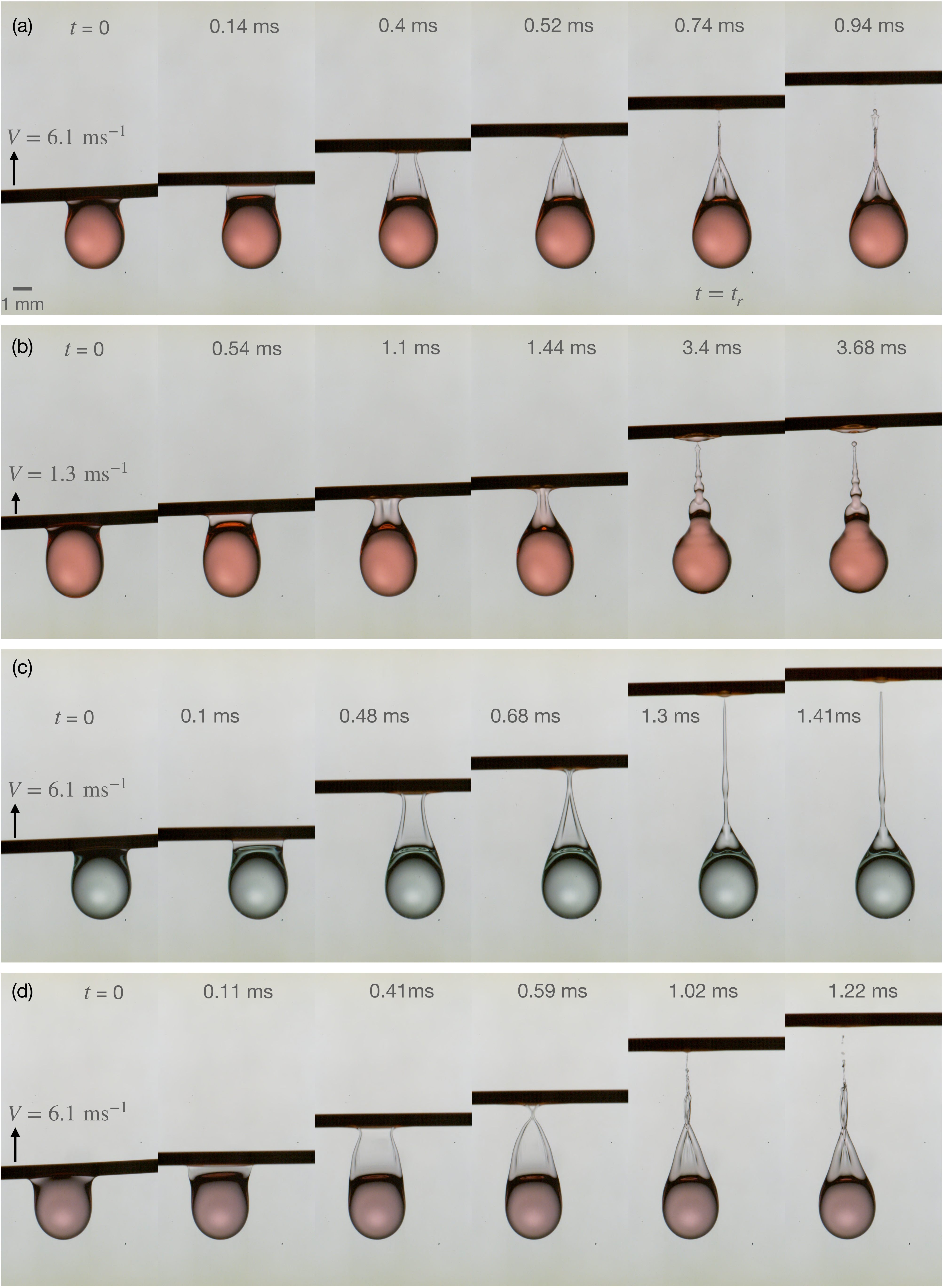}
\vspace{2mm}
\refstepcounter{figure}
\noindent\parbox{\textwidth}{%
{Figure \thefigure.}
{
Ejection dynamics of droplets for different fluids and wire speed on a wire of diameter $d_{\mathrm{w}}=$ 0.7 mm. The horizontal panels show temporal evolution for:  
(a) Water droplet (16~$\mu$l) with wire vertical speed $V = 6.1$ $\mathrm{ms^{-1}}$;  
(b) Water droplet (16~$\mu$l) with $V = 1.3$ $\mathrm{ms^{-1}}$;  
(c) Glycerin–water mixture droplet ($\mu = $25 mPa-s, 16~$\mu$l) with $V = 6.1$ $\mathrm{ms^{-1}}$;  
(d) Aqueous SDS droplet ($\sigma = $33 mN/m, 13~$\mu$l) with $V = 6.1$ $\mathrm{ms^{-1}}$ See supplementary videos 2-5}
}
\label{fig: dynamics}
\end{figure}

When a capillary droplet is placed on a horizontal wire, it adopts one of two characteristic shapes: a symmetric barrel shape or an asymmetric clamshell shape (\cite{lorenceau2004capturing, mei2013gravitational,  zhang2025dynamics}). In our experiments, gravity consistently renders the droplet asymmetric clamshell shape, resembling the hanging droplet  seen in Fig.\ref{fig: dynamics} at time $t=0$.

\begin{figure*}
\centering  
\includegraphics[width=1.0\textwidth]{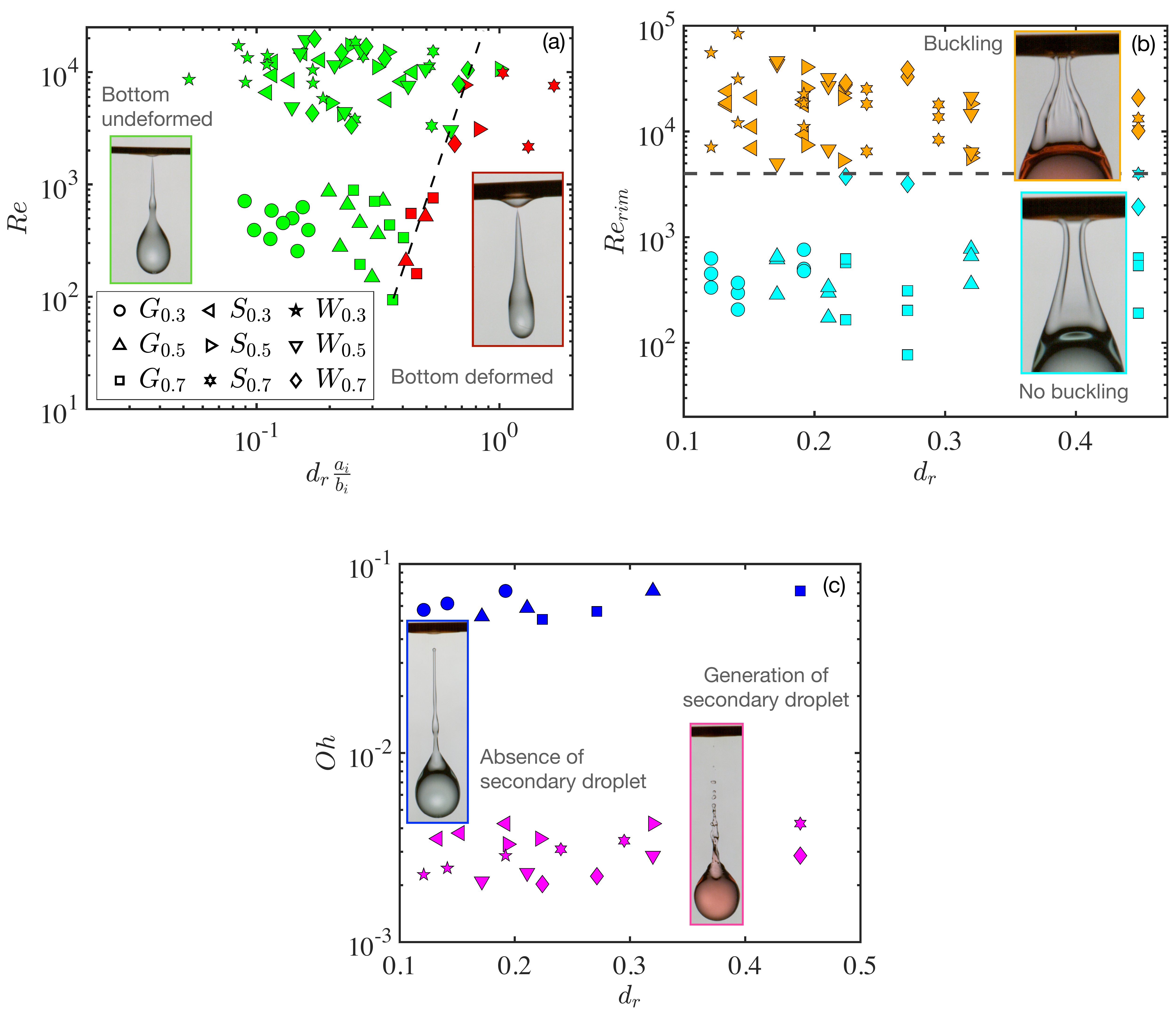}
\vspace{2mm}
\refstepcounter{figure}
\noindent\parbox{\textwidth}{%
{Figure \thefigure.}
{
(a) A regime map ($Re$ vs.\ $d_ra_i / b_i$) illustrates the extent of bottom deformation. Red markers indicated bottom deformation $>15\%$ and green markers $<15\%$, with two representative images indicating the distinct regimes and a dotted line marking their approximate boundary. Markers labeled $W$, $G$, and $S$ denote data for water, glycerin-water, and aqueous SDS droplets, respectively, with subscripts indicating the wire diameter.
(b) A regime map (rim Reynolds number ($Re_{\mathrm{rim}}$) vs. $d_r$) illustrates the onset of buckling on the liquid sheet. Two representative images depict the regimes, separated by a dotted boundary of $Re_{rim}=4000$.
(c) A regime map ($Oh$ vs. $d_r$) illustrates the onset of secondary droplet formation during jet stretching.
Two representative images depict the regimes.
}}
\label{fig: regimes}
\end{figure*}

In Fig.~\ref{fig: dynamics}a, a 16~$\mu$L water droplet with a spherical equivalent diameter of $d_{eq} = 3.1$~mm is initially placed on a copper wire of diameter $d_{\mathrm{w}} = 0.7$~mm at time $t = 0$.
 Upon plucking the wire, the lower portion of the droplet remains stationary, while the upper section, still in contact with the wire, is pulled upward with a velocity $V = 6.1$ $\mathrm{ms^{-1}}$.
 This motion elongates the droplet into a thin liquid sheet enclosed by two rims ($t =$~0.14 ms). 
 At $t = 0.4$~ms, a capillary wave emerges (see also
the top inset of Fig.~\ref{fig: regimes} b) on the sheet as the rims at the lateral edges migrate inward due to surface tension during the stretching process, similar to the buckling instability seen in \cite{marston2016crown}. These rims eventually collide and coalesce, giving rise to a single vertical jet at $t = 0.52$~ms.
The resulting jet continues to elongate vertically, becoming progressively thinner, and a chain of small fluid beads can be observed forming along the jet. Eventually, the jet detaches from the wire with a maximum vertical height $b_f$ at the release time $t_r = 0.74$~ms (column 6 of Fig. 2 lists the release time for all cases, which will be discussed later).
 Following detachment, the jet undergoes capillary-driven fragmentation, producing a cascade of smaller secondary droplets at $t = 0.94$~ms (see supplementary video 2). This sequence of events from initial sheet formation to rim collision and eventual jet breakup, highlights the central role of inertial and capillary forces in the breakup dynamics, and is consistent with phenomena reported in previous studies of ligament and sheet fragmentation (\cite{eggers1993universal,eggers1997nonlinear,wagner2003role, villermaux2004ligament, villermaux2009single,bhat2010formation, keshavarz2016ligament}).

In Fig.~\ref{fig: dynamics}b, we examined the dynamics under the same initial conditions, but with a reduced wire velocity of $V = 1.3~$ $\mathrm{ms^{-1}}$. The resulting behavior differs markedly from that observed at higher speeds. Even at $t=0$ Fig.~\ref{fig: dynamics}a and b are slightly different due to a slow vertical movement of the wire that occurs just before the high-speed impulsive motion due to the burning of the nylon support and wire tension, as carefully discussed in Fig.~S1 (i.e., the speed of retraction). 
However, at this lower velocity (Fig.~\ref{fig: dynamics}b), a significantly thicker fluid sheet and rim are formed. Consequently, the jet itself appears more voluminous, leading to the formation of larger fluid beads. The process takes considerably more time (higher $t_r$) to detach from the wire, highlighting the critical role of wire velocity in controlling the breakup and release dynamics of the fluid. The bottom of the droplet exhibits noticeable deformation compared to the case with higher wire velocity (see supplementary video 3). Although secondary droplets still form, their number is significantly lower than in the high-speed case.

We replaced the water droplet with a 25~mPa-s glycerin-water mixture while keeping the high-speed configuration ($V = 6.1$ $\mathrm{ms^{-1}}$) unchanged. In Fig.~\ref{fig: dynamics}c, we present the temporal evolution of the droplet shape at various time instants. Although the overall dynamics resemble those observed with pure water, the increased viscosity of the glycerin-water mixture causes a noticeable delay (higher $t_r$) in the detachment of the droplet from the wire and greater detachment length ($b_f$) of the droplet compared to water at that speed. The resulting jet exhibits a significantly smoother appearance, with no visible buckling on the sheet indicating enhanced stability during elongation and suppression of interfacial instabilities owing to the fluid higher viscosity (see supplementary video 4). Additionally, no secondary droplet generation is observed in this case.

To investigate the influence of surfactant on droplet detachment dynamics, we replaced the water droplet with a saturated aqueous SDS (sodium dodecyl sulfate) solution, having a surface tension of $\sigma = 33$~mN/m and fluid volume $13 \mu$l, while keeping the viscosity constant. Fig.\ref{fig: dynamics}d illustrates the temporal evolution of the droplet shape at various time instants. At this reduced surface tension, the static droplet shape at $t = 0$ differs markedly from that of water. The contact length ($a_i$) is longer because the contact angle is smaller, which influences both the rim collapse dynamics and the position of collapse ($t=0.59$~ms). Here, the rims collapse well below the wire, altering the breakup pathway and promoting pronounced buckling instabilities along with increased secondary droplet generation, likely due to the lowered surface tension and relatively higher $a_i$. Here, both the release time $t_r$ and the final droplet height $b_f$ are larger compared to those of water at the same wire speed (see supplementary video 5). Some interesting behaviors in the low–surface-tension cases, including breakup from the middle and vertical sheet formation, are discussed in Supplementary Section S2.

During the stretching process, the bottom of some droplets deform significantly. We investigate the onset of bottom deformation during the stretching phase 
by plotting the Reynolds number ($Re = \rho V {d_{eq}} / \mu$) versus $d_r a_i / b_i$, where $d_r = d_{\mathrm{w}} / d_{eq}$ in Fig.~\ref{fig: regimes}a.
Data points are color-coded based on deformation magnitude of a 15\% threshold (see Supplementary Section III for calculations). A dotted line demarcates the two regimes. 
Significant bottom deformation occurs at higher $d_ra_i/b_i$. 
The critical value of $d_ra_i/b_i$ increases with Reynolds number, implying that stronger inertia shifts the onset of base deformation to more confined geometries. 
At low $Re$, even moderate confinement can trigger bottom deformation, whereas at higher $Re$ the droplet resists base distortion until $d_ra_i/b_i$ becomes larger. 
This trend shows that the transition is not purely geometric but arises from the coupled competition between inertial forcing and confinement. 
Thus, inertia and geometric confinement together govern the onset of droplet base deformation.

To rationalize the onset of capillary waves, which correspond to an out-of-plane buckling of the sheet that emerges as stretching intensifies, 
we introduce a rim Reynolds number $Re_{rim} = \rho V_r d_{\mathrm{eq}} / \mu$, which measures the inertial loading
imposed by the moving rim relative to the viscous resistance of the sheet, where $V_r$ is the rim speed at the wire–water interface. In Fig.~\ref{fig: regimes}b, we plot $Re_{rim}$ against $d_r$ and find that the buckling systematically emerges once $Re_{rim} > 4000$. The threshold indicates that, beyond this point, the inertial forcing of the rim builds in-plane compressive stresses that exceed the effective compressive 
sheet strength, rendering the flat state unstable and triggering buckling. To our knowledge, buckling in such transient sheets generated by impulsive wire motion has not been previously examined.

The formation and stretching of the liquid jet can sometimes result in breakup and formation of secondary droplets. 
In Fig.~\ref{fig: regimes}c, we found that the Ohnesorge number 
$Oh = \mu/\sqrt{\rho \sigma d_{eq}}$
plays a central role in determining which jets form secondary droplets and which do not. We chose to use $d_{eq}$ in the calculation of $Oh$ because it is a more simple and accurate measurement, whereas the radius of the jet changes with time and distance from the wire. However, if we were to use the average radius of the jet at breakup instead of the equivalent diameter the trend would remain though the values in the figure would be approximately three times larger.   At low \(Oh\) the surface tension dominates over viscosity and the jet becomes increasingly unstable, leading to the formation of small secondary droplets. 
Conversely, at higher \(Oh\) (for instance, in glycerin-water mixture jets), viscous effects dominate, resulting in smoother jets with suppressed secondary droplet formation. Representative experimental images for each regime are shown in the insets to highlight the visual contrast between them. 

Detachment occurs when the droplet stretching reaches a maximum height $b_f$ at the release time $t_r$, and we observe that $b_f$ varies systematically across fluids and experimental conditions. Because detachment occurs at this point of maximum extension, predicting the vertical stretch $\Delta z = b_f - b_i$ provides a minimal description of the entire impulsive event (i.e., $t_r = \Delta z / V$). We therefore seek a scaling for $\Delta z$ in terms of the measurable input parameters. During the interval of nearly constant upward wire motion, the wire transmits an interfacial force to the droplet through the three-phase contact line. The mechanical work performed on the droplet up to release is therefore set by this force acting over the vertical stretch. This input work is partly stored as an increase in capillary energy associated with the enlarged free-surface area, while the remainder is dissipated by viscous deformation within the elongating filament.

\begin{figure}
\centering
\includegraphics[width=0.55\textwidth]{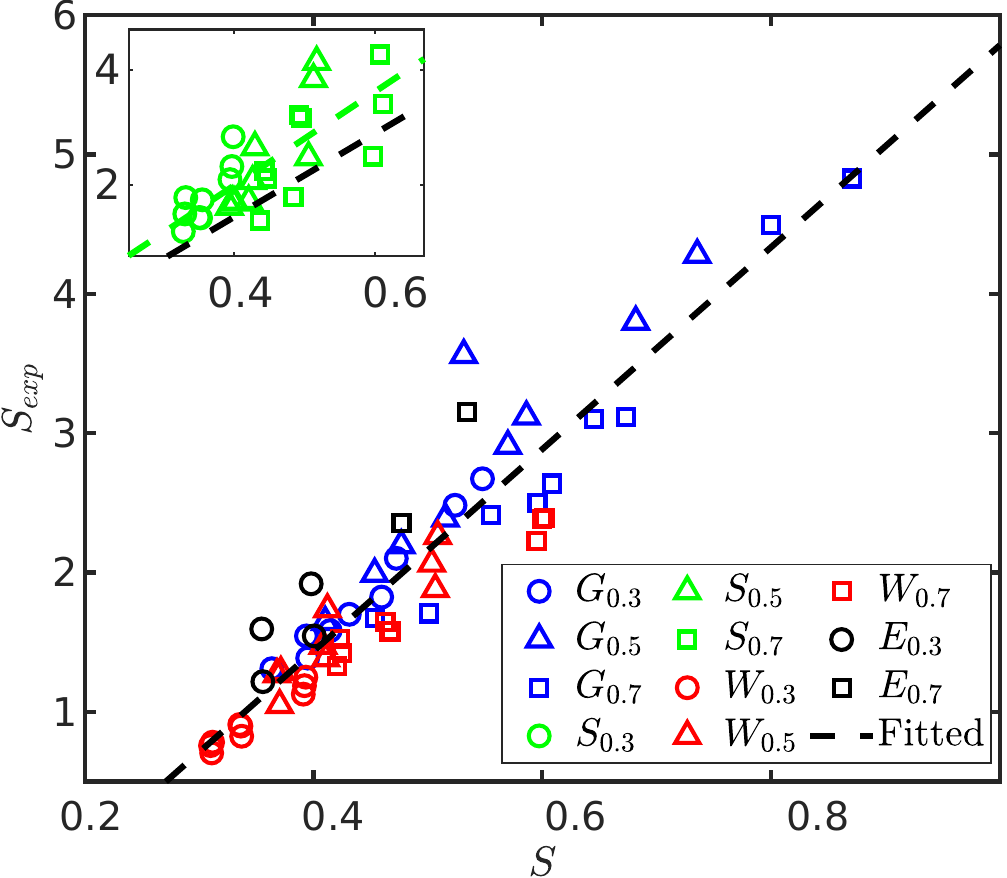}
\vspace{2mm}
\refstepcounter{figure}
\noindent\parbox{\textwidth}{%
{Figure \thefigure.}
{
Plot of predicted non-dimensional stretching $S$ versus experimentally measured non-dimensional stretching $S_{exp}$. 
Red, blue, green, and black markers denote water, glycerin--water mixtures, aqueous SDS solutions, ethanol, respectively. 
The black dashed line shows the best fit to the water, glycerin--water, and ethanol data while the green dashed line 
shows the best fit to the SDS data.
Markers labeled $W$, $G$, $S$, $E$ denote data for water, glycerin-water, and aqueous SDS droplets, and pure ethanol, respectively, with subscripts indicating the wire diameter.}}
\label{fig: scaling}
\end{figure}

At the wire--droplet contact, the upward traction is transmitted through the
three-phase contact line. We therefore estimate the available upward force as
the surface-tension force integrated along the contact-line length ($L_{\mathrm{cl}}$),
$F_{\mathrm{up}} = \sigma\,L_{\mathrm{cl}}
= \sigma\left(\pi d_w + 2  d_{eq}\right).$
The first contribution represents the azimuthal contact line around
the wire, approximated as one-half of the wire circumference, and we take the
contact angle to be zero for simplicity of the model. The second contribution accounts for the axial extent of the
contact region along the wire, where $d_{eq}$ sets the droplet length scale. This estimate captures the leading-order capillary traction transmitted at the three-phase line during rapid stretching; corrections enter only as geometric prefactors and do not alter the functional dependence.
Therefore, the work done during stretching over the vertical extension $\Delta z$
scales as 
\begin{equation}
W_{\mathrm{up}} = \sigma\left(\pi d_w + 2 d_{eq}\right)  \Delta z
\end{equation}




Stretching increases the interfacial curvature and therefore generates an
excess Laplace pressure $\delta P$. For the cone-like interface sketched in
Fig.~\ref{Fig:schematic}(e,f), we write the Laplace pressure (\cite{landau1987fluid}) as
$\delta P \sim \sigma\!\left(\partial^2 y/\partial z^2+\partial^2 x/\partial z^2\right)$.
Here $y$ is $a_i$, which is of order $d_{eq}$, and $x$ is of order $\pi d_{\mathrm{w}}/2$,
with variations over $\Delta z$. We therefore estimate
$\partial^2 y/\partial z^2 \sim d_{eq}/(\Delta z)^2$ and
$\partial^2 x/\partial z^2 \sim \pi d_{\mathrm{w}}/2(\Delta z)^2$. The associated
capillary work scale is set by this excess pressure acting over a
characteristic area $\pi d_{eq} d_{\mathrm{w}}/2$ over the vertical extension $\Delta z$.
\begin{equation}
W_{c} = \frac{\sigma (\pi d_{\mathrm{w}} d_{eq})(\pi d_{\mathrm{w}}/2+d_{eq})}{2{\Delta z}}.
\end{equation}

Viscous dissipation energy over the vertical length up to stretching is 
\begin{equation}
W_{\mu} = \frac{\mu V \pi d_{\mathrm{w}} d_{eq}}{2}.
\end{equation}
Defining the non-dimensional stretching as $S = \Delta z/d_{\mathrm{eq}}$, introducing the dimensionless group $d_r = d_{\mathrm{w}}/d_{\mathrm{eq}}$,  capillary number 
($Ca = \mu V/\sigma$), and applying the energy balance ($W_{\mathrm{up}}=W_{\sigma}+W_{\mu}$) yields
\begin{equation}
S \;=\; \frac{\pi d_r\,\mathrm{Ca}}{4(\pi d_r+2)}
\left[
1
+
\sqrt{
1
+
\frac{4(\pi d_r+2)^2}{\pi d_r\,\mathrm{Ca}^2}
}
\right].
\label{eq:S}
\end{equation}

In Fig.~\ref{fig: scaling}, the measured non-dimensional stretching $S_{exp}$ is compared with the prediction of eq.\ref{eq:S}, which depends only on $Ca$ and $d_r$. Across droplet volumes, wire diameters, and pure fluids (water, glycerin-water, ethanol), the data collapse onto a single approximately linear trend.  The model therefore captures the dominant scaling of stretching during the interval of constant upward wire motion. The SDS solutions also follow the same slope, however, the time-dependent interfacial tension ($\sigma_{\text{SDS}}(t)$) that SDS solutions are known to exhibit during rapid deformation
(\cite{speirs2018entry}), results in modified capillary traction at the contact line. 


The agreement in Fig.\ref{fig: scaling} with eq.\ref{eq:S} reveals that the visually rich breakup sequence is ultimately constrained by a simple energetic balance during the stretching stage. Although the subsequent dynamics involve sheet formation, rim collision, and jet fragmentation, the maximum extension of the droplet before release is governed by capillary traction transmitted through the three-phase contact line and resisted by viscous dissipation and curvature growth. When expressed in terms of the capillary number $Ca$ and geometric ratio $d_r$, the stretching collapses across droplet sizes, wire diameters, and clean fluids onto a single organizing curve. In this sense, the plucked-wire event reduces to a minimal balance between contact-line forcing and interfacial restoration and viscous dissipation. The deviation observed for SDS solutions identifies the regime where additional interfacial transport and Marangoni stresses can actively modify this traction, marking the boundary of validity of the clean-interface description.


\section{Conclusion}
In conclusion, impulsive detachment of droplets from fibers can be understood as a capillary-mediated transmission of mechanical impulse into interfacial deformation. While the breakup pathway exhibits multiple observable regimes: bottom deformation, sheet buckling, and secondary droplet generation; the extent and duration of stretching prior to release are governed by a compact energetic law depending only on $Ca$ and $d_r$. This reduction condenses a complex transient process into a low-dimensional organizing principle for impulsively forced droplets. By identifying when this collapse holds and when additional interfacial physics intervenes, the plucked-wire configuration provides a controlled platform for probing how rapid forcing couples to capillary interfaces, with implications for fiber-based droplet shedding (\cite{challita2023droplet,hoggarth2025producing}) and manipulation in both natural and engineered systems.



\backsection[Supplementary data]{\label{SupMat}Supplementary material and movies are available at \\https://doi.org/xx.xxxx/jfm.xxxx...}



\backsection[Declaration of interests]{ The authors report no conflict of interest.}






\bibliographystyle{jfm}
\bibliography{jfm}

@article{zhang2025dynamics,
  title={Dynamics of inviscid droplets on fibres: from instability to oscillations},
  author={Zhang, Fei and Zhao, Shuguang and Zhou, Xinping},
  journal={Journal of Fluid Mechanics},
  volume={1011},
  pages={A34},
  year={2025},
  publisher={Cambridge University Press}
}

@article{lohse2022fundamental,
  title={Fundamental fluid dynamics challenges in inkjet printing},
  author={Lohse, Detlef},
  journal={Annual review of fluid mechanics},
  volume={54},
  number={1},
  pages={349--382},
  year={2022},
  publisher={Annual Reviews}
}

@article{villermaux2009single,
  title={Single-drop fragmentation determines size distribution of raindrops},
  author={Villermaux, Emmanuel and Bossa, Benjamin},
  journal={Nature physics},
  volume={5},
  number={9},
  pages={697--702},
  year={2009},
  publisher={Nature Publishing Group UK London}
}

@article{gilet2025leaf,
  title={Leaf oscillation and upward ejection of droplets in response to drop impact},
  author={Gilet, Tristan and Tadrist, Lo{\"\i}c},
  journal={Physical Review Fluids},
  volume={10},
  number={5},
  pages={053601},
  year={2025},
  publisher={APS}
}

@article{lejeune2019drop,
  title={Drop impact close to the edge of an inclined substrate: liquid sheet formation and breakup},
  author={Lejeune, Sophie and Gilet, Tristan},
  journal={Physical Review Fluids},
  volume={4},
  number={5},
  pages={053601},
  year={2019},
  publisher={APS}
}

@article{pan2020regime,
  title={Regime map and triple point in selective withdrawal},
  author={Pan, Zehao and Nunes, Janine K and Stone, Howard A},
  journal={Physical review letters},
  volume={125},
  number={26},
  pages={264502},
  year={2020},
  publisher={APS}
}

@article{vincent2014remnants,
  title={Remnants from fast liquid withdrawal},
  author={Vincent, Lionel and Duchemin, Laurent and Villermaux, Emmanuel},
  journal={Physics of Fluids},
  volume={26},
  number={3},
  year={2014},
  publisher={AIP Publishing}
}

@article{eggers1993universal,
  title={Universal pinching of 3D axisymmetric free-surface flow},
  author={Eggers, Jens},
  journal={Physical review letters},
  volume={71},
  number={21},
  pages={3458},
  year={1993},
  publisher={APS}
}

@article{eggers1997nonlinear,
  title={Nonlinear dynamics and breakup of free-surface flows},
  author={Eggers, Jens},
  journal={Reviews of modern physics},
  volume={69},
  number={3},
  pages={865},
  year={1997},
  publisher={APS}
}

@article{bhat2010formation,
  title={Formation of beads-on-a-string structures during break-up of viscoelastic filaments},
  author={Bhat, Pradeep P and Appathurai, Santosh and Harris, Michael T and Pasquali, Matteo and McKinley, Gareth H and Basaran, Osman A},
  journal={Nature Physics},
  volume={6},
  number={8},
  pages={625--631},
  year={2010},
  publisher={Nature Publishing Group UK London}
}

@article{villermaux2004ligament,
  title={Ligament-mediated spray formation},
  author={Villermaux, Emmanuel and Marmottant, Ph and Duplat, J{\'e}rome},
  journal={Physical review letters},
  volume={92},
  number={7},
  pages={074501},
  year={2004},
  publisher={APS}
}

@article{keshavarz2016ligament,
  title={Ligament mediated fragmentation of viscoelastic liquids},
  author={Keshavarz, Bavand and Houze, Eric C and Moore, John R and Koerner, Michael R and McKinley, Gareth H},
  journal={Physical review letters},
  volume={117},
  number={15},
  pages={154502},
  year={2016},
  publisher={APS}
}

@article{wagner2003role,
  title={Role of inertia in two-dimensional deformation and breakdown of a droplet},
  author={Wagner, AJ and Wilson, LM and Cates, ME},
  journal={Physical review E},
  volume={68},
  number={4},
  pages={045301},
  year={2003},
  publisher={APS}
}

@article{bai2011controlled,
  title={Controlled fabrication and water collection ability of bioinspired artificial spider silks},
  author={Bai, Hao and Ju, Jie and Sun, Ruize and Chen, Yuan and Zheng, Yongmei and Jiang, Lei},
  journal={Advanced Materials},
  volume={23},
  number={32},
  pages={3708--3711},
  year={2011},
  publisher={Wiley Online Library}
}

@article{ju2012multi,
  title={A multi-structural and multi-functional integrated fog collection system in cactus},
  author={Ju, Jie and Bai, Hao and Zheng, Yongmei and Zhao, Tianyi and Fang, Ruochen and Jiang, Lei},
  journal={Nature communications},
  volume={3},
  number={1},
  pages={1247},
  year={2012},
  publisher={Nature Publishing Group UK London}
}

@article{marston2016crown,
  title={Crown sealing and buckling instability during water entry of spheres},
  author={Marston, Jeremy O and Truscott, Tadd T and Speirs, Nathan B and Mansoor, Mohammad M and Thoroddsen, Sigurdur T},
  journal={Journal of Fluid Mechanics},
  volume={794},
  pages={506--529},
  year={2016},
  publisher={Cambridge University Press}
}

@article{dickerson2012wet,
  title={Wet mammals shake at tuned frequencies to dry},
  author={Dickerson, Andrew K and Mills, Zachary G and Hu, David L},
  journal={Journal of the Royal Society Interface},
  volume={9},
  number={77},
  pages={3208--3218},
  year={2012},
  publisher={The Royal Society}
}

@article{ortega2012aerial,
  title={Aerial shaking performance of wet {A}nna's hummingbirds},
  author={Ortega-Jimenez, Victor Manuel and Dudley, Robert},
  journal={Journal of The Royal Society Interface},
  volume={9},
  number={70},
  pages={1093--1099},
  year={2012},
  publisher={The Royal Society}
}

@article{wang2025bioinspired,
  title={Bioinspired fibers with alternating wettability and elastic vibration: enhancing water harvesting},
  author={Wang, Xikui and Luo, Hong and Wei, Han and Zhou, Xueqiu and Qin, Bingli and Mei, Yi and Zhang, Youfa},
  journal={Separation and Purification Technology},
  pages={133180},
  year={2025},
  publisher={Elsevier}
}

@article{mei2013gravitational,
  title={The gravitational effect on the geometric profiles of droplets on horizontal fibers},
  author={Mei, Maofei and Fan, Jintu and Shou, Dahua},
  journal={Soft Matter},
  volume={9},
  number={43},
  pages={10324--10334},
  year={2013},
  publisher={Royal Society of Chemistry}
}

@article{lorenceau2004capturing,
  title={Capturing drops with a thin fiber},
  author={Lorenceau, {\'E}lise and Clanet, Christophe and Qu{\'e}r{\'e}, David},
  journal={Journal of colloid and interface science},
  volume={279},
  number={1},
  pages={192--197},
  year={2004},
  publisher={Elsevier}
}

@book{landau1987fluid,
  title={Fluid Mechanics},
  author={Landau, Lev Davidovich and Lifshitz, Evgeny Mikhailovich},
  volume={6},
  year={1987},
  publisher={Elsevier}
}

@article{speirs2018entry,
  title={Entry of a sphere into a water-surfactant mixture and the effect of a bubble layer},
  author={Speirs, NB and Mansoor, Mohammad M and Hurd, Randy Craig and Sharker, Saberul I and Robinson, WG and Williams, BJ and Truscott, Tadd T},
  journal={Physical Review Fluids},
  volume={3},
  number={10},
  pages={104004},
  year={2018},
  publisher={APS}
}

@article{farhan2018universal,
  title={Universal expression for droplet--fiber detachment force},
  author={Farhan, Noor M and Vahedi Tafreshi, H},
  journal={Journal of Applied Physics},
  volume={124},
  number={7},
  year={2018},
  publisher={AIP Publishing}
}

@article{ojaghlou2018dynamical,
  title={Dynamical insights into the mechanism of a droplet detachment from a fiber},
  author={Ojaghlou, Neda and Tafreshi, Hooman V and Bratko, Dusan and Luzar, Alenka},
  journal={Soft matter},
  volume={14},
  number={44},
  pages={8924--8934},
  year={2018},
  publisher={Royal Society of Chemistry}
}

@article{challita2023droplet,
  title={Droplet superpropulsion in an energetically constrained insect},
  author={Challita, Elio J and Sehgal, Prateek and Krugner, Rodrigo and Bhamla, M Saad},
  journal={Nature communications},
  volume={14},
  number={1},
  pages={860},
  year={2023},
  publisher={Nature Publishing Group UK London}
}

@inproceedings{hoggarth2025producing,
  title={Producing Droplets on a Vibrating Liquid Bath Through Ligament Stretching and Breakup},
  author={Hoggarth, Johnathan and Harris, Daniel M and Bush, John WM and Primkulov, Bauyrzhan K},
  booktitle={Division of Fluid Dynamics Annual Meeting 2025},
  year={2025},
  organization={APS}
}

\end{document}


\maketitle

\begin{center}
{ Supplementary Materials}

\vspace{5pt}
\rule{3in}{0.5pt}
\vspace{8pt}
\end{center}
\begin{figure}
\centering
\includegraphics[width=0.65\linewidth]{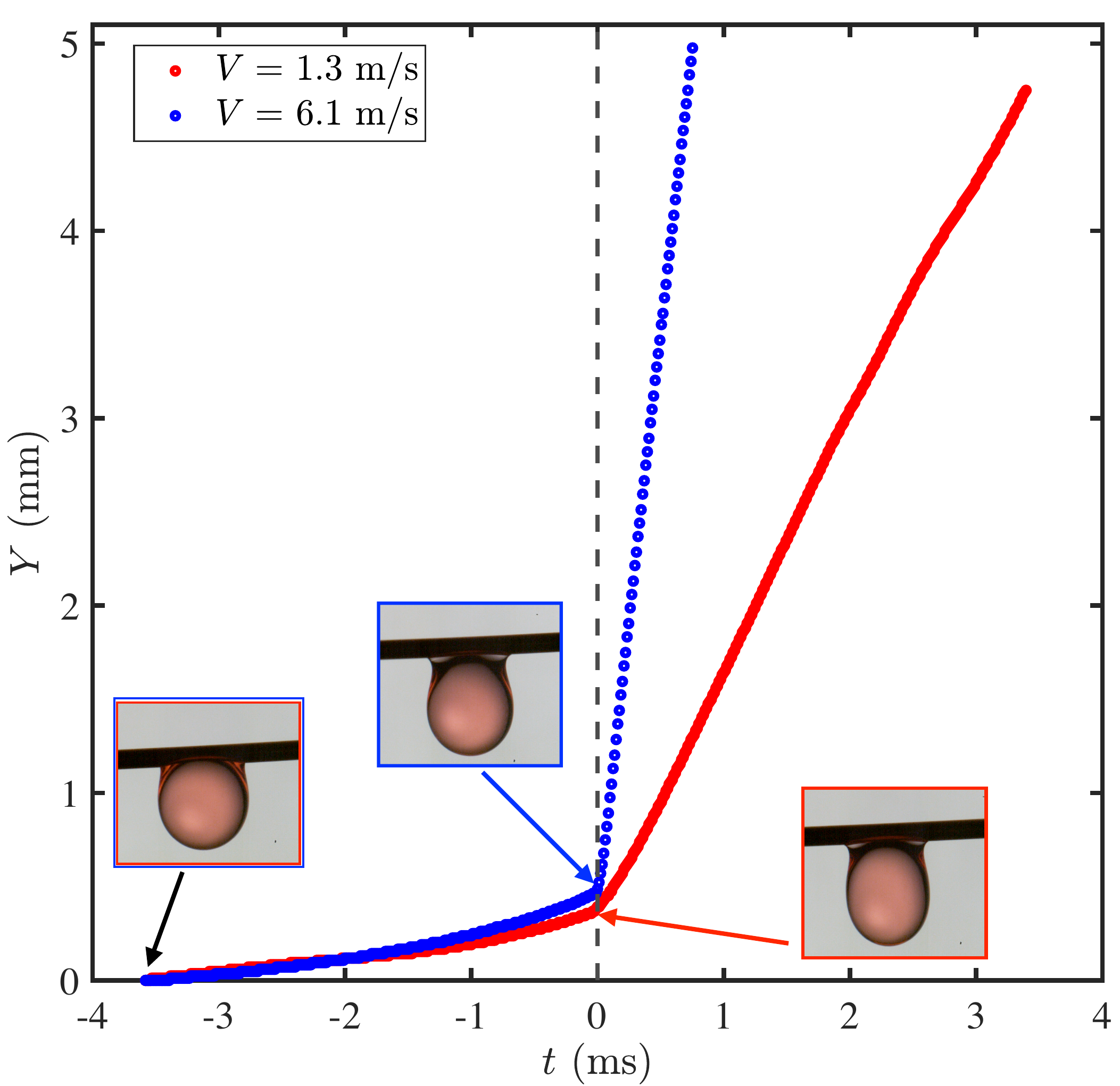} 
\vspace{2mm}
\refstepcounter{figure}
\noindent\parbox{\textwidth}{%
{Figure S\thefigure.}
{Vertical position of the wire as a function of time for two different wire
speeds $V$, with $d_{\mathrm{w}} = 0.7~\mathrm{mm}$. The experimental images with arrows indicate the shape of a water droplet with
$d_{\mathrm{eq}} = 3.1~\mathrm{mm}$ at that time for two different wire speeds.
}}
\end{figure}

\begin{figure}
\centering
\includegraphics[width=0.4\linewidth]{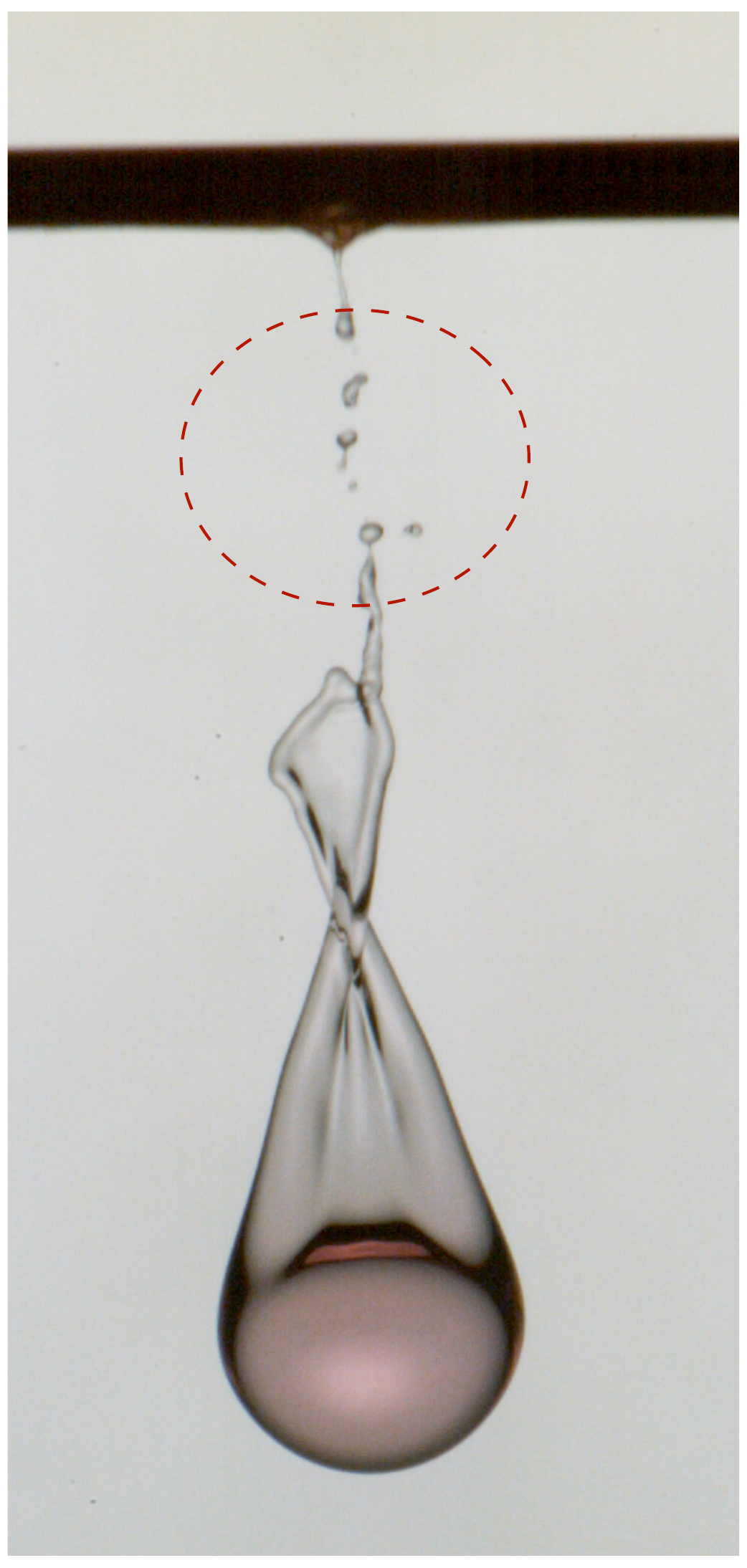} 
\vspace{2mm}
\refstepcounter{figure}
\noindent\parbox{\textwidth}{%
{Figure S\thefigure.}
{A jet emerges from an aqueous SDS droplet during stretching and breaks up near the middle for 
$d_{\mathrm{w}} = 0.7~\mathrm{mm}$, $d_{\mathrm{eq}} = 2.4~\mathrm{mm}$, and 
$V = 6.1~\mathrm{m\,s^{-1}}$.
}}
\end{figure}

\begin{figure}
\centering
\includegraphics[width=0.65\linewidth]{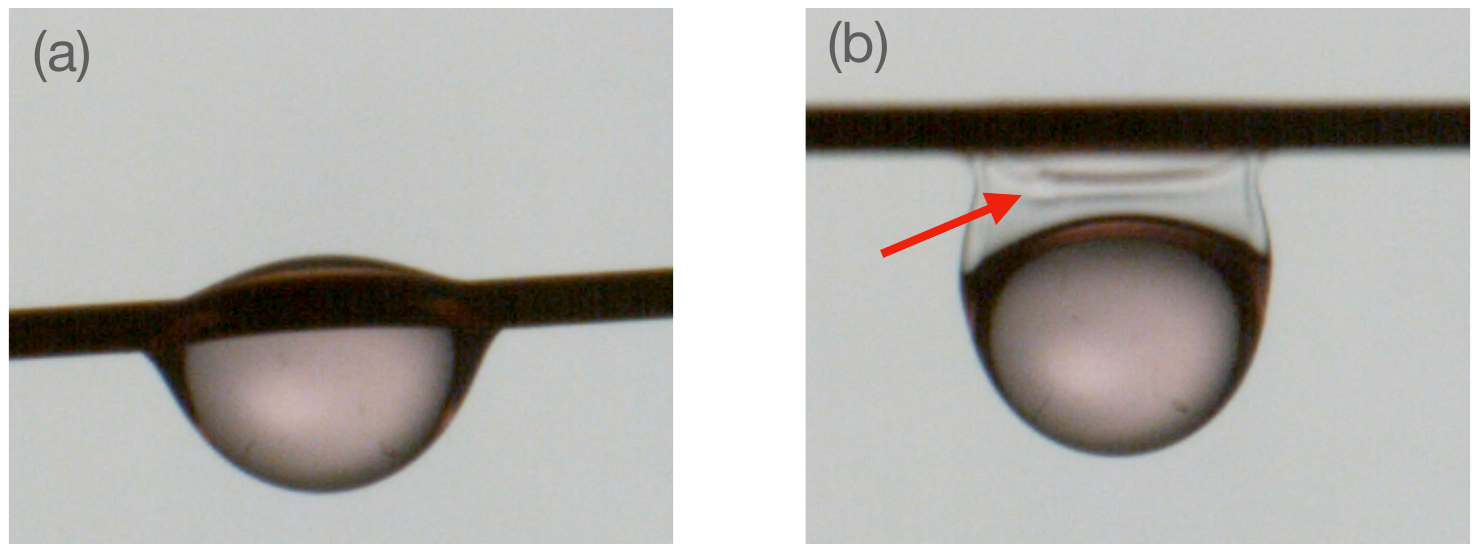} 
\vspace{2mm}
\refstepcounter{figure}
\noindent\parbox{\textwidth}{%
{Figure S\thefigure.}
{Emergence of vertical buckling in the liquid sheet of an aqueous SDS droplet
during stretching for
$d_{\mathrm{w}} = 0.3~\mathrm{mm}$, $d_{\mathrm{eq}} = 1.6~\mathrm{mm}$, and
$V = 6.1 ~\mathrm{m\,s^{-1}}$.
}}
\end{figure}

\begin{figure}
\centering
\includegraphics[width=1.0\linewidth]{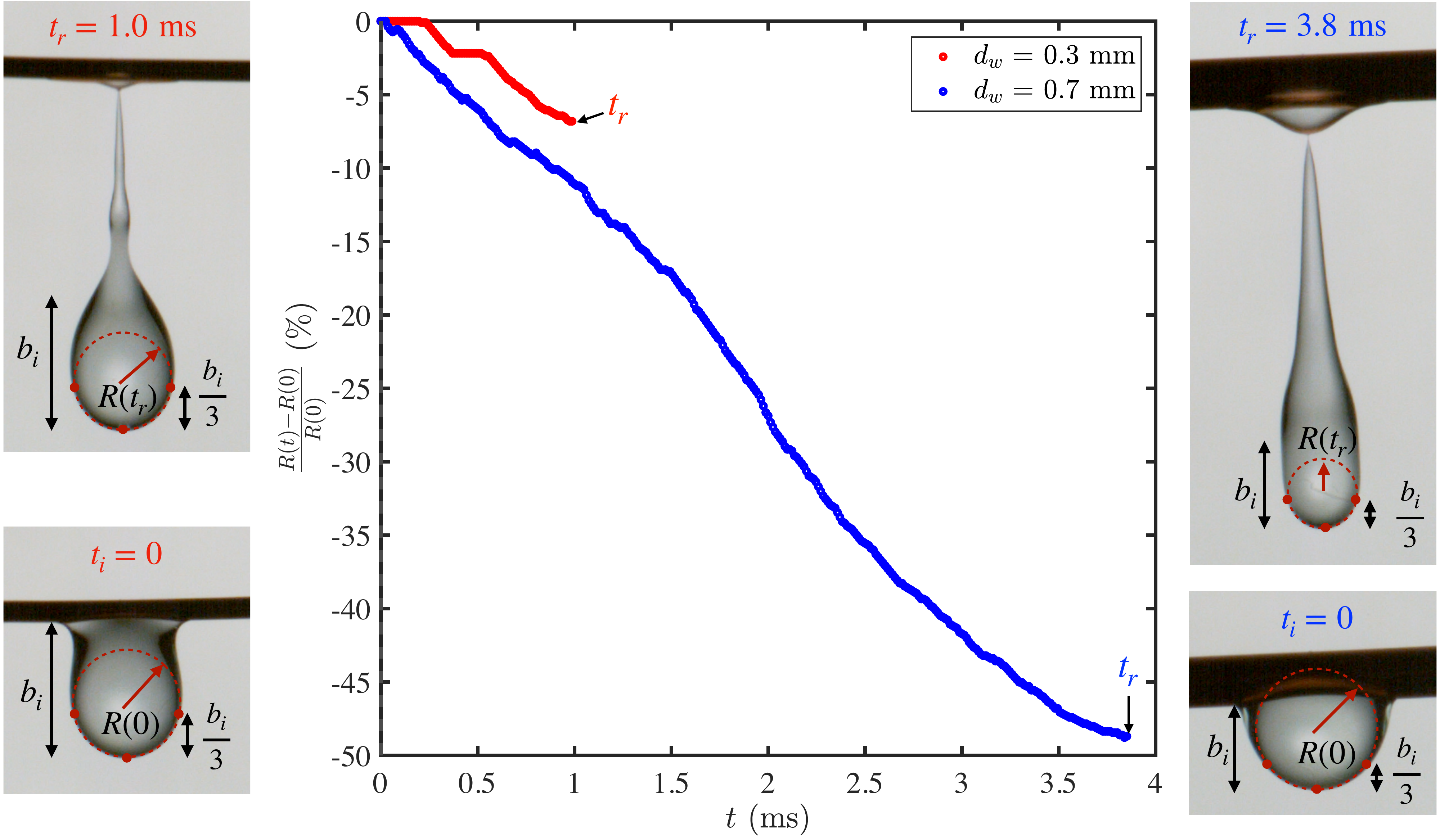} 
\vspace{2mm}
\refstepcounter{figure}
\noindent\parbox{\textwidth}{%
{Figure S\thefigure.}
{ 
Variation of the normalized bottom radius, $\bigl[R(t)-R(0)\bigr]/R(0)$, is plotted for a glycerin-water droplet with $d_{\mathrm{eq}} = 1.6~\mathrm{mm}$ for two wire conditions with $d_{\mathrm{w}} = 0.3~\mathrm{mm}$ and $V = 3.2~\mathrm{m\,s^{-1}}$ and with $d_{\mathrm{w}} = 0.7~\mathrm{mm}$ and $V = 1.3~\mathrm{m\,s^{-1}}$ in the central plot. The experimental images on the left correspond to the first case and those on the right correspond to the second case, and they show the droplet shape at the initial and release time.}}
\end{figure}

\textbf{S1 Droplet's initial shape}

To ensure a well-defined initial condition, our experimental system is carefully optimized to achieve a stable base state prior to the application of the impulsive excitation. The impulse is triggered by burning a nylon wire; however, this ignition process is not truly instantaneous and introduces a finite delay before the full impulse is delivered. We define $t = 0$ as the moment corresponding to the onset of the sharp change in velocity, marking the initiation of the impulse as shown in Fig.S1. Nonetheless, residual droplet motion is observed even at negative times ($t < 0$), which arises due to pre-impulse dynamics influenced by the tension in the wire. Specifically, the red and blue curves correspond to two different tension (2.94~N and 53.96~N) conditions associated with $d_{\mathrm{w}} = 0.7$~mm. These differing tensions result in distinct wave speeds and consequently yield different initial droplet shapes at $t = 0$. The experimental images marked with arrows depict the shape of a water droplet with $d_{\mathrm{eq}} = 3.1~\mathrm{mm}$ at the corresponding instants for the two wire speeds.
\\\\
\textbf{S2 Jet breaks up from the middle and vertical buckling}

In some cases of aqueous SDS droplets, jet detachment occurs along the jet rather than at the wire–droplet interface, as shown in Fig.S2. This premature breakup is attributed to the lower surface tension, which facilitates early rupture along the stretched filament. Additionally, in a few instances, vertical buckling of the fluid sheet was observed, as shown in Fig.S3. This behavior appears to originate from the initial droplet shape, where a small volume of fluid remains above the wire even before motion begins. This residual fluid, remaining at rest while the wire moves upward, seeds vertical buckling during the subsequent stretching phase.
\\\\
\textbf{S3 Bottom deformation calculation}

To differentiate the onset of bottom deformation in the droplet, we plotted the Reynolds number ($Re$) against the ratio $d_ra_i/b_i$, as shown in Fig.~3a. The deformation was quantified by estimating the radius of curvature at the bottom of the droplet through a three-point circle fitting method, as shown by the red dot in the experimental images of Fig~S4. To choose the three points, we selected the lowest point of the droplet and the leftmost and rightmost points from the bottom at a vertical distance of $b_i/3$ from the very bottom, where $b_i$ is the maximum vertical height of the droplet at $t = 0$. Using these points, we computed the instantaneous radius of curvature $R(t)$ until the release time $t_r$ and plotted the normalized variation of it, $(R(t) - R(0))/R(0)$, in Fig.S4. We found that the variation at $t_r$ was minimal, less than $15\%$, for a glycerin–water droplet with $d_{\mathrm{eq}} = 1.6~\mathrm{mm}$, $d_{\mathrm{w}} = 0.3~\mathrm{mm}$, and $V = 3.2~\mathrm{m\,s^{-1}}$, whereas for $d_{\mathrm{w}} = 0.7~\mathrm{mm}$ and $V = 1.3~\mathrm{m\,s^{-1}}$ the variation was approximately $48\%$, indicating significant deformation. Based on this analysis, we defined a $15\%$ change in curvature as the cutoff threshold that separates bottom-undeformed and bottom-deformed cases, as shown in Fig.~3a.


\newpage
\textbf{S4 SM Video captions}\\
\textbf{Video 1.} High-speed motion of raindrops flung from grass blades following a sudden movement of the grass. Compare with Fig.~1(a).

\textbf{Video 2.} Ejection dynamics of 16~$\mu\mathrm{L}$ droplets of water on a wire of diameter $d_{\mathrm{w}} = 0.7~\mathrm{mm}$ with $V = 6.1~\mathrm{m\,s^{-1}}$. Compare with Fig.~2(a).

\textbf{Video 3.} Ejection dynamics of 16~$\mu\mathrm{L}$ water droplets on a wire of diameter $d_{\mathrm{w}} = 0.7~\mathrm{mm}$ for $V = 1.3~\mathrm{m\,s^{-1}}$. Compare with Fig.~2(b).

\textbf{Video 4.} Ejection dynamics of 16~$\mu\mathrm{L}$ droplets of glycerin–water mixture on a wire of diameter $d_{\mathrm{w}} = 0.7~\mathrm{mm}$ with $V = 6.1~\mathrm{m\,s^{-1}}$. Compare with Fig.~2(c).

\textbf{Video 5.} Ejection dynamics of 13~$\mu\mathrm{L}$ droplets aqueous SDS solution on a wire of diameter $d_{\mathrm{w}} = 0.7~\mathrm{mm}$ with $V = 6.1~\mathrm{m\,s^{-1}}$. Compare with Fig.~2(d).

\textbf{Video 6.} Glycerin-water mixture droplet dynamics with $d_{\mathrm{eq}} = 1.6~\mathrm{mm}$ for $d_{\mathrm{w}} = 0.3~\mathrm{mm}$ and $V = 3.2~\mathrm{m\,s^{-1}}$. Compare with Fig.~3(a).

\textbf{Video 7.} Glycerin-water mixture droplet dynamics with $d_{\mathrm{eq}} = 1.6~\mathrm{mm}$ for  $d_{\mathrm{w}} = 0.7~\mathrm{mm}$ and $V = 1.3~\mathrm{m\,s^{-1}}$. Compare with Fig.~3(a).

\textbf{Video 8.} Water droplet dynamics with $d_{\mathrm{eq}} = 3.1~\mathrm{mm}$, $d_{\mathrm{w}} = 0.7~\mathrm{mm}$ and $V = 4.5~\mathrm{m\,s^{-1}}$. Compare with Fig.~3(b).

\textbf{Video 9.} Glycerin-water mixture droplet dynamics with $d_{\mathrm{eq}} = 3.1~\mathrm{mm}$, $d_{\mathrm{w}} = 0.7~\mathrm{mm}$ and $V = 4.5~\mathrm{m\,s^{-1}}$. Compare with Fig.~3(b).

\textbf{Video 10.} Water droplet dynamics with $d_{\mathrm{eq}} = 3.1~\mathrm{mm}$, $d_{\mathrm{w}} = 0.7~\mathrm{mm}$ and $V = 4.5~\mathrm{m\,s^{-1}}$. Compare with Fig.~3(c).

\textbf{Video 11.} Glycerin-water mixture droplet dynamics with $d_{\mathrm{eq}} = 3.1~\mathrm{mm}$, $d_{\mathrm{w}} = 0.7~\mathrm{mm}$ and $V = 4.5~\mathrm{m\,s^{-1}}$. Compare with Fig.~3(c).

\textbf{Video 12.} Temporal evolution of the droplet shape of a 16~$\mu\mathrm{L}$ water droplets with wire diameter $d_{\mathrm{w}} = 0.7~\mathrm{mm}$, for wire velocities $V = 1.3~\mathrm{m\,s^{-1}}$ during burning of the nylon wire. Compare with Fig.~S.1.

\textbf{Video 13.} Temporal evolution of the droplet shape of a 16~$\mu\mathrm{L}$ water droplets with wire diameter $d_{\mathrm{w}} = 0.7~\mathrm{mm}$, for wire velocities  $V = 6.1~\mathrm{m\,s^{-1}}$ during burning of the nylon wire. Compare with Fig.~S.1.

\textbf{Video 14.} Compare with Fig.~S.2. 

\textbf{Video 15.} Compare with Fig.~S.3. 
